\documentclass[pre,twocolumn,nofootinbib,showpacs,superscriptaddress,10pt]{revtex4-1}

\pdfoutput=1
\usepackage[english]{babel}
\usepackage[utf8]{inputenc}
\usepackage{graphicx}
\usepackage{hyperref}
\usepackage{gensymb}
\usepackage{setspace}
\usepackage{amsmath}
\begin{document}
\title{Clustering of bacteria with heterogeneous motility}

\author{T. Vourc'h}
\altaffiliation{Present address: Physico-Chimie Curie, CNRS, Institut Curie, Paris Sciences et Lettres, 11 rue Pierre et Marie Curie 75005 Paris, France}
\affiliation{Laboratoire AstroParticules et Cosmologie, CNRS, Universit\'e Paris-Diderot, Universit\'e de Paris, 5 rue Thomas Mann 75013 Paris, France}

\author{J. L\'eopold\`es}
\email{julien.leopoldes@espci.fr}
\affiliation{ESPCI Paris, PSL Research University, CNRS, Institut Langevin, 1 rue Jussieu, F-75005 Paris, France}
\affiliation{Universit\'e Paris-Est Marne-la-Vall\'ee, 5 Bd Descartes, Champs sur Marne, Marne-la-Vall\'ee Cedex 2, France}

\author{H. Peerhossaini}
\affiliation{Laboratoire AstroParticules et Cosmologie, CNRS, Universit\'e Paris-Diderot, Universit\'e de Paris, 5 rue Thomas Mann 75013 Paris, France}
\affiliation{Mechanics of Active Fluids Laboratory, Department of Civil and Environmental Engineering, Department of Mechanical and Materials Engineering, University of Western Ontario, London, ON N6A 3K7, Canada}
\date{\today}
\begin{abstract}
\par We study the clustering of a model cyanobacterium \textit{Synechocystis} into microcolonies. The bacteria are allowed to diffuse onto surfaces of different hardness, and interact with the others by aggregation and detachment. We find that soft surfaces give rise to more microcolonies than hard ones. This effect is related to the amount of heterogeneity of bacteria's dynamics as given by the proportion of motile cells. A kinetic model that emphasizes specific interactions between cells, complemented by extensive numerical simulations considering various amounts of motility, describes the experimental results adequately. The high proportion of motile cells enhances dispersion rather than aggregation.
\end{abstract}
\keywords{cyanobacteria, biofilm, motility, microcolonies}

\maketitle

\section*{Introduction}
Divided matter tends to cluster on a wide range of lengthscales, depending on the range of the driving force and the type of interaction. At the molecular level, proteins aggregate in neurons as a possible mechanism for some neurodegenerative diseases~\cite{neuro}. Moreover, the clustering of mesoscopic phytoplankton allows vertical transport in the oceans~\cite{marine} while at largest scales, the asteroid families in the main asteroid belt are believed to result from the aggregation that follows a collision~\cite{accr}.

Distinct from the above examples, motion in active matter such as bacteria is driven by non-equilibrium forces. The aggregation between bacterial cells in microcolonies (clusters), a fundamental step in the colonisation of a surface, triggers the formation of a biofilm. This favors the adaptation of bacteria to their local environment by constituting a protective system against external toxic agents~\cite{Kumar2017,Thurlow2011,Bjarnsholt2013,planchon,Landry2006,Alhede2011}. Biofilm formation is a major concern in healthcare or food industry, but its control could also be profitable for decontamination or renewable energies. 

The patterns observed following the aggregation of the cells are different from inert particles~\cite{benjacob}. Indeed, the morphology of the biofilm reflects the complex conditions under which growth occurs, such as gradients of nutriment or light as well as number density and motility~\cite{Klausen2006}. Nevertheless, some similarities exist between active cluster formation and first order phase transitions in thermal systems~\cite{catestailleur,baskaran}. 

Focusing on the dynamics at the particle scale, the balance between nucleation-division and diffusion-aggregation processes shall control the emergence of microcolonies~\cite{Duvernoy2018,Weber2015,Bonazzi2018,
Taktikos2015}. For example, motility can either favor bacterial aggregation by enabling cell-cell encounters~\cite{Weber2015} but also prevent localized aggregates by enhancing dispersion~\cite{Acemel2018}. Moreover, it is well-known that bacteria develop several subpopulations in order to adapt to evolving environmental conditions~\cite{Smits2006}, so that heterogeneous behaviours can be found inside colonies~\cite{Oldewurtel2015,Ponisch2018}. But how heterogeneity influences clustering in such real systems is presently unknown. 

In this article, we study the growth of microcolonies of cyanobacteria Synechocystis on soft and hard surfaces. The latter promote higher amounts of motile bacteria $p_m$ than the former and additionally, the number of clusters at long times is a decreasing function of $p_m$. We propose that motility allows the bacteria to escape from clusters while non-motile ones are trapped. This study highlights the necessity to account for subpopulations of variable dynamics among a given strain for an adequate description of the formation of microcolonies. 
\begin{figure*}[t]
\centering
\includegraphics[scale=0.4]{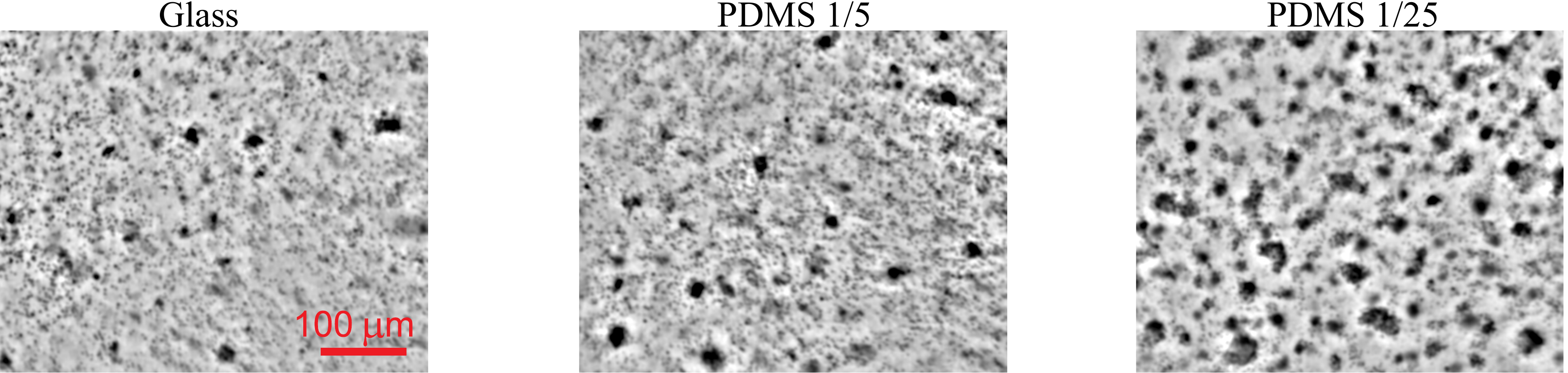}
\caption{\label{Fig1} Morphology of the biofilm on different substrates, 8 days after introduction of the bacteria in the cell. Glass, PDMS 1/5 and PDMS 1/25 correspond to hard, intermediate and softer surface, respectively.}
\end{figure*}

\section*{Experiments}
Biofilms of the wild-type strain of the cyanobacterium \textit{Synechocystis} sp. PCC 6803 are grown during eight days on surfaces of various hardness fixed at the bottom of the same Petri dish (``open cell"), allowing identical gas exchange and exposure to the controlled light. After introduction of the bacterial suspension in the open cell, the latter is placed into an incubator for biofilm growth. Once a day, the morphology of the various samples is imaged with a microscope. Typically, a raw image is filtered with a bandpass filter to reduce noise, before subtraction of the background, inversion and binarisation. The remaining image contains a distribution of areas of contiguous pixels $\mathcal{A}_i$ which are selected based on the size of their equivalent radius $r_i=\sqrt{\mathcal{A}_i/\pi}$ compared to the radius of the bacteria $r_b$. All areas with $r_i<1.4$~$\mu$m, corresponding to a lower bound for the size of a Synechocystis, are removed from the analysis. The number of individual bacteria contained in the rest of each area is obtained from the floor function of $r_i/r_b$ ($r_b=2.2$~$\mu$m) which is then summed over all the areas to compute $\rho(t)$. The number of areas with $r_i>3.1$~$\mu$m defines the number of microcolonies.

We also characterized the surface specific dynamics of bacteria in a ``closed cell" with no gas exchange and designed to prevent bacteria motion due to convection~\cite{Lenotre}. In the closed cell, the use of dilute suspensions at short time scales limits the interactions between bacteria and allows the study of their individual diffusive motion, mediated by type IV pili. Video recording one hour after introduction of the suspension in the closed cell and subsequent particle tracking provides the measurement of the ``run" and ``tumble" times of the intermittent diffusion, characteristic of this system~\cite{Lenotre}. No microcolony forms in the closed cell as the density remains low and the duration of the experiment is much less than the division time.
The experiments in either open or closed cells have been achieved with glass, PDMS 1/5 and PDMS 1/25 as a substrate, corresponding to hard, intermediate and soft surfaces of respective Young modulus of $5.10^3$, $3.6$ and $1$ MPa. Details on the culture protocol and on the materials are provided in the Additional information.

\section*{Results}
\begin{figure}[h]
\centering
\includegraphics[scale=0.38]{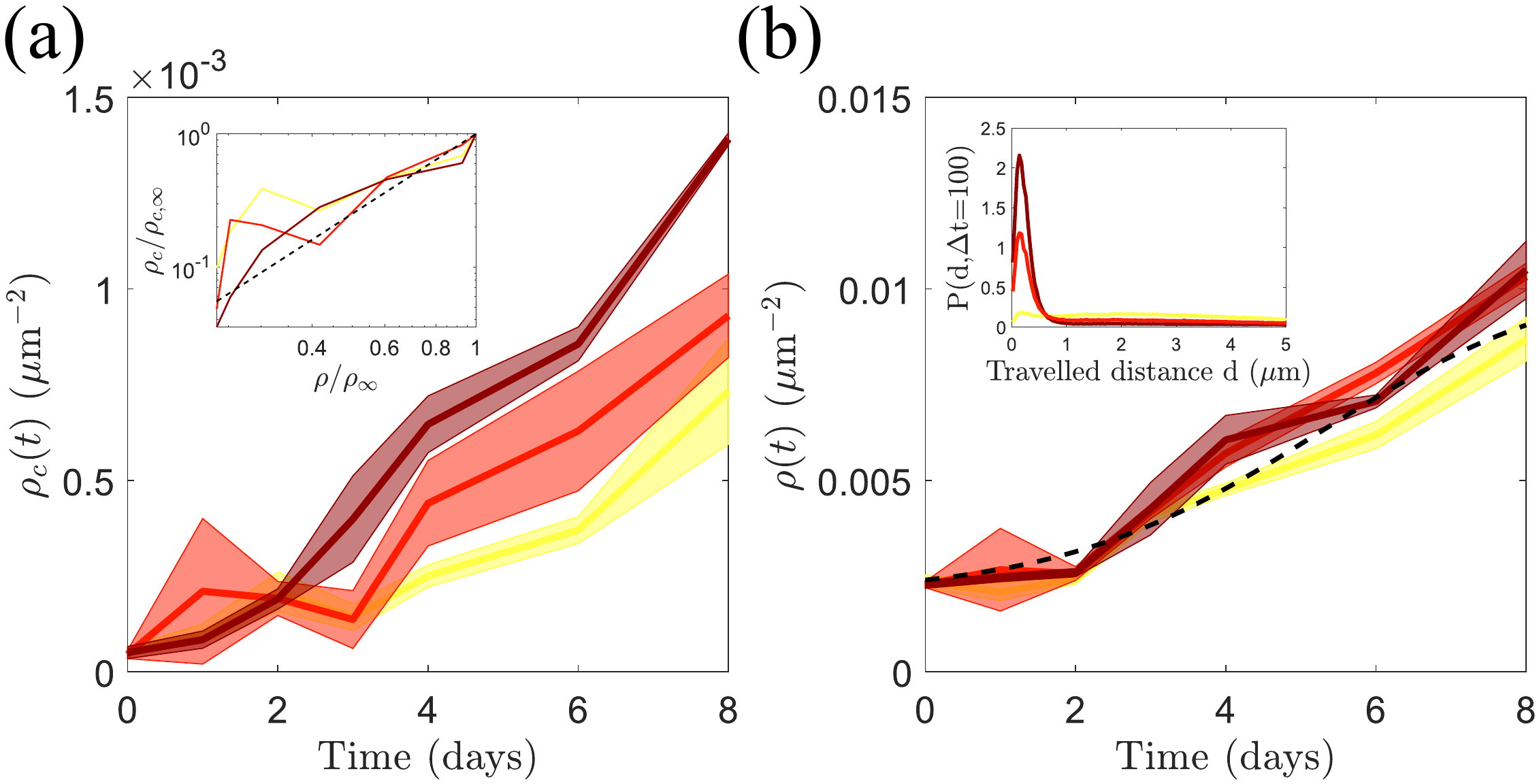}
\caption{\label{Fig2} (a) Number density of microcolonies as a function of time. From the darkest to the brightest line: PDMS 1/25, 1/5 and glass. Inset: logarithmic plot of the number of microcolonies as a function of $\rho$, with both axis rescaled by the final experimental value. Black dashed line indicates a slope 2. (b) Number of detected bacteria on the surface. Plain line: experimental results. Black dashed line: Logistic growth model, see text. Inset: probability density function of displacement of individual bacteria obtained in the closed cell. Color code identical for (a) and (b).}
\end{figure}

Figure~\ref{Fig1} shows some micrographs taken in the open cell on the different surfaces after eight days of growth, corresponding to the final time of the experiment of biofilm growth. The large black spots consist of microcolonies that result from the clustering of bacteria. The microcolonies contain a small number of cells, typically 10 at maximum. Smaller black dots are individual bacteria. Clearly, the soft surface PDMS 1/25 is covered by numerous clusters while fewer are detected on the hard glass. Quantification of this effect is reported in Figure~\ref{Fig2}(a) where the surface number density of microcolonies is plotted as a function of time. At early times all surfaces are covered by the same number of microcolonies that formed in suspension before the start of the experiment. At larger times, the number of microcolonies is nearly three times as large on the soft PDMS 1/25 than on the hard glass one. 

The kinetics of cellular growth is, however, not dependent on the type of substrate as shown in Figure~\ref{Fig2}(b) where the surface number density of bacteria $\rho(t)$ is plotted against time. We adjust $\rho(t)$ with the logistic equation $\rho(t)=
\rho_i+\frac{\rho_\infty-\rho_i}{1+[(\rho_\infty-\rho_i)/\rho_i-1]exp(-(t-\lambda)/\tau_{div})}$, where $\rho_i=2.10^{-3} \mu$m$^{-2}$ and $\rho_\infty=11.10^{-3} \mu$m$^{-2}$ are the initial and the final number density of bacteria~\cite{Zwietering1990}. $\lambda=3.2\pm0.4$ days defines a period of latency such that $\rho(\lambda)=2\rho_i$ and $\tau_{div}$=40 $\pm7$ hours is the division time.

For a given $\rho$, the various phenotypes shown Fig.~\ref{Fig1} should arise from specific surface dynamics, which we have studied in details in the closed cell. The probability distribution functions of displacements (time interval $\Delta t=100$ s), shown in Figure~\ref{Fig2}(b) (inset), reveal marked peaks at distances smaller than the radius of the bacteria (lower bound $r_a~\sim 1.4~\mu$m), indicating the presence of immobile cells. The broad tails characterize motile ones.

Analysis of the trajectories of motile bacteria on each type of surface (closed cell) shows that they evolve diffusively, with the ubiquitous intermittent ``run'' and ``tumble'' dynamics associated with identical caracteristic times $\langle \tau_{r}\rangle \sim 5$~s and $\langle \tau_{t}\rangle \sim 15$~s, leading to a diffusion coefficient $D \sim$ 0.05 $\mu$m$^2$.s$^{-1}$ as in~\cite{Lenotre}. Motile bacteria have the same dynamics whatever the substrate. 

Figure~\ref{Fig2}(b) (inset) suggests that soft surfaces give rise to a higher number of unmotile bacteria $n_u$, as shown by the amplitudes of the peaks at small displacements. $n_u$ is computed by tracking cells that do not travel a distance larger than $2$ $\mu$m during $\Delta t=100$~s~$>~\langle \tau_{t}+\tau_r\rangle$. The resulting proportion of motile bacteria $p_m=n_m/(n_m+n_u)$ amounts to $p_m = 0.9$, $0.7$ and $0.25$ for the glass, PDMS 1/5 and 1/25 respectively. This leads to our main experimental result: microcolonies form preferentially on softer surfaces, corresponding to a low proportion of motile cells.
\paragraph*{Modeling and numerical simulations}
Experiments show that the clusters are not motile. Thus, we consider that attachment-detachment events involve only one single bacterium and a cluster. For simplicity, we disregard cell growth and consider homogeneous population for now. Let $c_i$ be the surface number density of the clusters containing $i$ cells and $J_i$ the net rate at which the $i$ clusters transform to $i+1$ ones. Following~\cite{Penrose1989} we have $\dot{c_i}=J_{i-1}-J_{i}$ for $i\geq 2$ and $\dot{c_1}=-2J_{1}-\sum_{i=2}^\infty J_{i}$. The $J_i$ are given by equations $J_i=a_ic_1c_i-b_{i+1}c_{i+1}$, where $a_i$ and $b_i$ are kinetic coefficients ($[a]=L^2T^{-1}$ and $[b]=T^{-1}$). This infinite system of differential equations forms the Becker-D\"{o}ring discrete equations for fragmentation-coagulation process with conservation of mass $\rho=\sum_{i\geq 1}ic_i=n/S=const$~\cite{Penrose1989,Becker1935}. 
\begin{figure}[h]
\centering
\includegraphics[scale=0.3]{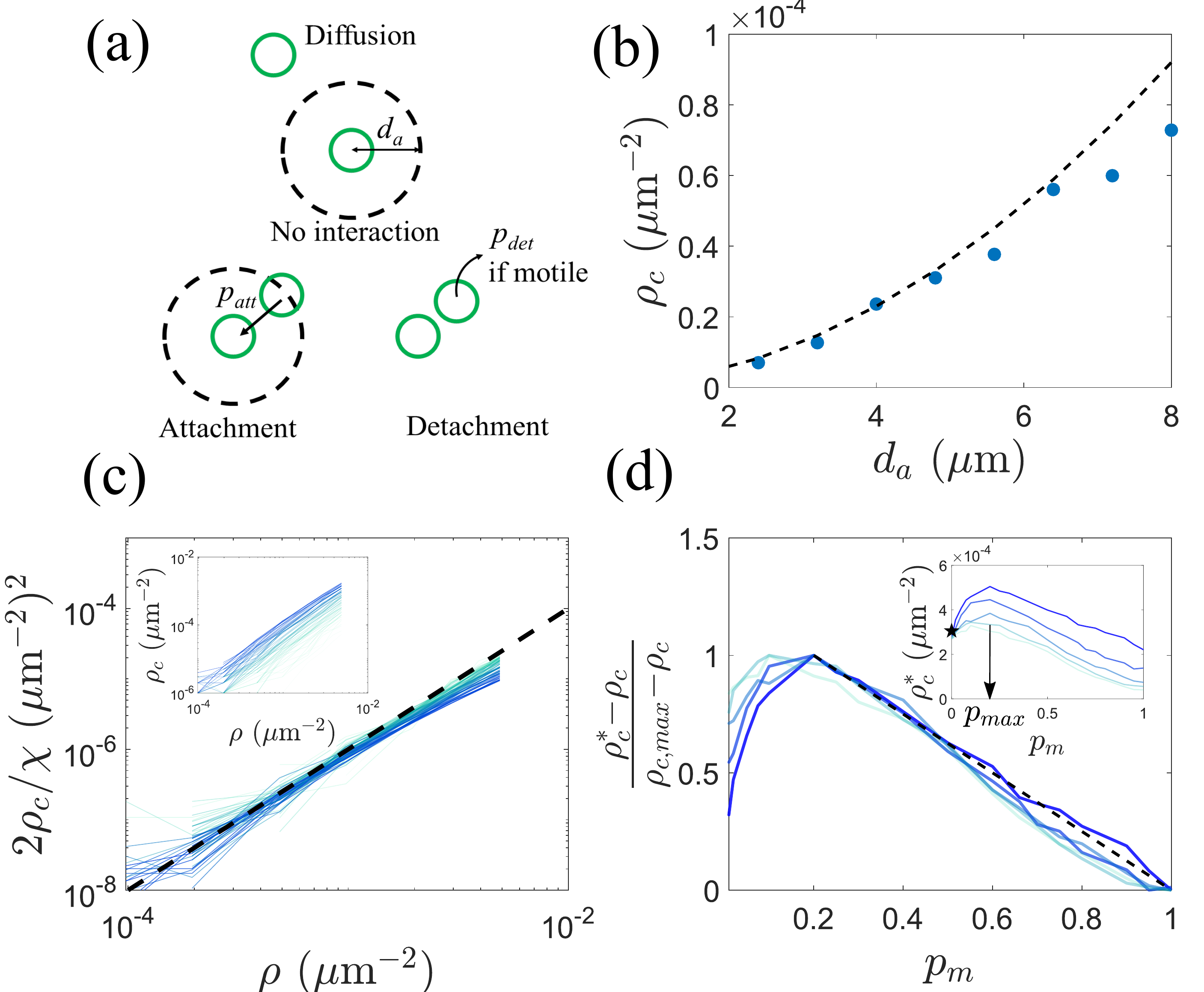}
\caption{\label{Fig3} Numerical simulations. (a) Sketch of the mechanism for attachment and detachment processes. (b) Number density of microcolonies versus $d_a$. Dashed line represents the model Eq.~\ref{becker3}. Each points is the average of 10 runs. (c) Numerical results rescaled with Eq.~\ref{becker3} (dashed line indicates a slope 2). Parameters $100<t_a<1000$ s, $50<t_b<500$ s and $2 <d_a <8$ $\mu$m. Each curve is averaged over four runs. Darker curves correspond to higher values of $\chi=\pi d_a^2 t_b/t_a$. Inset: $\rho_c$ as a function of $\rho$ before rescaling. (d) Inset: final number of microcolonies with heterogeneous motility $\rho_c^*$, as a function of $p_m$ (same color code as (c)). The star indicates theoretical predictions considering a Rayleigh-distributed interparticle distance (see text). Main graph: rescaled number of microcolonies as a function of $p_m$. The black dashed line indicates the fit by $(1-p_m)/(1-p_{max})$ for $p_m>p_{max}$.}
\end{figure}

\begin{figure*}[t]
\centering
\includegraphics[scale=0.4]{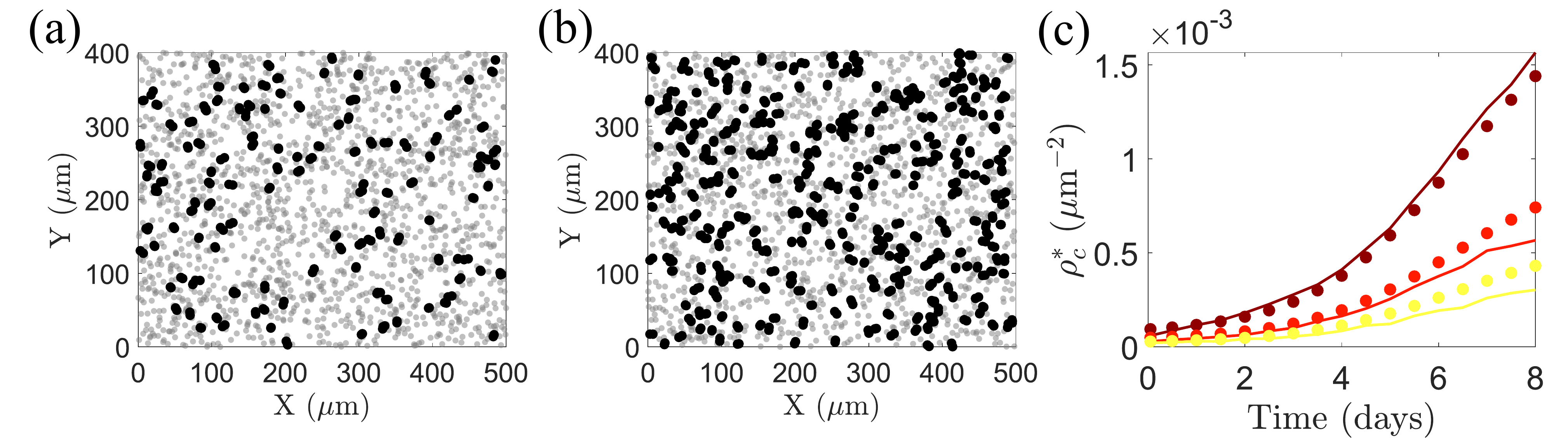}
\caption{\label{Fig4} Simulations with bacterial growth, run with $d_a$=3 $\mu$m $t_a=450$s and $t_b=100$s. (a) and (b): colonisation maps at long times, on a "high motility" and a "low motility" surface, defined with $p_m$=0.7 and 0.25, respectively. Unbound and/or motile cells are in light gray, unmotile and attached ones are in black. (c): Number of detected microcolonies as a function of time for $p_m$=0.9, 0.7 and 0.25. Lines: simulation results. Dots: Eq.~\ref{eqFin}. Darkest lines correspond to lowest motilies.}
\end{figure*}

Since the clusters contain only a few bacteria, we assume that the rate at which the cells bind to and are released from microcolonies is $a_i=i\times a$ and $b_i=i\times b$, respectively. We focus on the long time limit where $J_i=0$ ($i\geq 1$). Then, the solution of the recursive relations gives $c_i=\frac{1}{i}\left(\frac{a}{b}\right)^{i-1}\left(c_1\right)^i$ from which $c_1=\frac{\rho}{1+\chi \rho}$ where $\chi=a/b$. With $\sum_{i\geq 1}c_i=c_1+\rho_c$ ($\rho_c$ is the number density of clusters), we obtain $\rho_c=-\frac{1}{\chi} \ln(1-\chi c_1)-c_1$. In the limit $\chi \rho \ll 1$, the order of magnitude for $c_1 \sim \rho$ and combining the expression of $\rho_c$ Taylor-expanded to second order yields:
\begin{equation}
\label{becker3}
\rho_c \sim \frac{\chi}{2}\rho^2
\end{equation}
Let us emphasize that in this long time limit, $\rho_c$ will not depend on transport as given by the average diffusion coefficient. The characteristic time of the coagulation-fragmentation process $t^\star \sim 1/(b+a\rho)\sim t_b =100$ s (see below) provides a condition $\frac{\Delta n}{n} =\frac{t^\star}{n} \tau_{div}^{-1}\sim 10^{-6}\ll1$ for the number of cluster reaching equilibrium before significant cell growth has occurred. The quadratic form Equation~\ref{becker3} agrees satisfactorily with the results Figure~\ref{Fig2}(a) (inset), although the appearance of tridimensional growth after eight days of experiments limits the relevance of the analysis to a timescale too short for evidencing of a proper scaling.

Akin to inert particles~\cite{chandra}, we assume that each bacterium is surrounded by a "disk of influence" of radius $d_a$ such that every bacteria entering this disk has a finite probability to stick. However, the interaction of between bacteria is different from inert colloids and we assume that it is driven by (type IV) pili-pili attachment, as in other similar systems~\cite{Bonazzi2018,Oldewurtel2015}. Correspondingly, we consider that attachment and detachment events occur within well-defined characteristic timescales $t_a$ and $t_b$, respectively. Then, dimensional analysis suggests $a\sim \pi d_a^2/t_a$ and $b \sim 1/t_b$. We now test this, together with Eq.~\ref{becker3}, by numerical simulations. 

As sketched Figure~\ref{Fig3}(a), the bacteria in the numerical simulations diffuse and interact according to the following rules: (i) only single bacteria can be motile (in proportion $p_m$), following Brownian motion with an effective diffusion coefficient $D$ (ii) every particle entering the disk of influence of another one has a probability $p_{att}=\delta t / t_a$ to stick, where $\delta t$ is the simulation step (iii) every bound particle that can be motile (in proportion $p_m$) can detach with probability $p_{det}=\delta t / t_b$.

We start with the simplest case where the proportion of motile bacteria $p_m=1$, and take an order of magnitude for $t_b\sim t_a \sim 100$~s~\cite{Zaburdaev2014,Taktikos2015,Bonazzi2018}. As shown in Figure~\ref{Fig3}(b), the number density of microcolonies $\rho_c$ from simulations follows Eq.~\ref{becker3} when $d_a$ varies, with $\chi=a/b= \pi d_a^2 t_b /t_a$ and the restriction $R<d_a <1/\rho^{1/2}$ since the radius of the disk of influence shall be much smaller than the average inter-particle distance. Moreover, Figure~\ref{Fig3}(c) indicates that simulations match the scaling Eq.~\ref{becker3} for a broad range of particle densities and characteristic times $t_a$ and $t_b$.

Then, we plot on Figure~\ref{Fig3}(d) (inset) the number density of microcolonies $\rho_c^*$ for different levels of heterogeneity as obtained from various proportions of motile bacteria $p_m$. Numerical results agree with the experimental data qualitatively and less microcolonies are obtained with the high proportion of motile bacteria. When no particle is motile ($p_m$=0), the interaction is only possible if two cells are situated within the radius of influence $d_a$. Assuming that cells are located randomly on the surface, the distance to the closest neighbor is distributed with a Rayleigh law~\cite{Torquato1990}, with parameter $\sigma= 1/\sqrt{2\,\pi\,\rho}$. Accordingly, the cumulative distribution function gives $p_{d \leq d_a}=1-exp(-d_a^2/(2\sigma^2))$ and a number of pairs $\sim p_{d \leq d_a} \times \rho/2$ [star in the inset Figure~\ref{Fig3}(d)]. For $p_m=1$, $\rho_c^*=\rho_{c}$ as given by Eq.\ref{becker3} and shown in Fig.\ref{Fig3}(d). In between, $\rho_c^*$ reaches a maximum $\rho_{c,max}$ at a given motility $p_{max}\sim 0.2$ common to all simulations [Figure~\ref{Fig3}(d) (inset)]. Thorough numerical investigation of the maximum shows that $\rho_{c,max}-\rho_{c} \sim 1/2 \pi d_a^2 \rho^2$ and linear extrapolation between $\rho_{c,max}$ and $\rho_{c}$ [Fig.\ref{Fig3}(d)] provides the final result for the surface number density of microcolonies for different amounts of motility:
\begin{equation}
\label{eqFin}
\rho_c^* \sim \frac{1}{2}\,\pi d_a^2\, \left[\frac{t_b}{t_a}\, + \frac{1-p_m}{1-p_{max}} \right]\,\rho^2
\end{equation}
valid for $p_m > p_{max}$, relevant for our experiments. The first term in the brackets corresponds to Eq~\ref{becker3}, while the second one provides influence of motility. In Eq.~\ref{eqFin}, $t_a$, $t_b$ and $d_a$ describe cell/cell interactions, while $p_m$ stands for the degree of heterogeneity of the motility, whose relation with cell/substrate interactions for the present study is discussed below. Above the timescale $\tau= \lambda+\tau_{div}^{-1}\log(\mathcal{S}/(N_i\,\pi\,d_a^2)) \sim 8$ days, with $\mathcal{S}=2.10^5$ $\mu$m$^2$ the surface area onto which initially $N_i=400$ particles are present, we expect tridimensional growth to dominate biofilm growth. 

Data analysis consistently show that for $10 < \chi < 100$ $\mu$m$^2$, a variation of $p_m$ from 0.2 to 1 increases the ratio of attachment/detachment events of 20$\%$. While this effect deserves further theoretical investigations, this clearly highlights that enhanced motility favors the escape of bacteria from microcolonies. This is the main mechanism responsible for the lower number of microcolonies on hard surfaces.

Now, we set $d_a=2\,r_a\sim 3$~$\mu$m for the order of magnitude of the interaction length. A measure of the detachment time $t_b\sim 100$ s can be obtained from experiments by the analysis of the video recorded in the closed cell, by detecting some detachment events and counting the time elapsed between attachment and detachment of two individual bacteria. Using Eq.~\ref{eqFin} along with the experimental data for ($p_m$, $t_b$) and the estimate for $d_a$ mentionned above provides a consistent characteristic time for attachment $t_a=450 \pm 150$ s whatever the surface. This is compatible with the order of magnitude found for different bacteria with type IV pili~\cite{Taktikos2015,Bonazzi2018}, and could suggest an ubiquitous interaction mechanism for bacteria displaying hair-like appendages. Our model suggests that the nature of the substrate only affects motility and not cell-cell interactions. 

Finally, we show in Figure~\ref{Fig4}(a-b) the morphologies obtained at long times from full simulations on "high-motility" and "low-motility" surfaces ($p_m$=0.7 and $0.25$, respectively) where bacterial growth is taken into account as described by the Logistic equation Figure~\ref{Fig2}(b) (details in Additional info). Consistent with experiments, small and sparse microcolonies are obtained, with more clusters on the "low-motility" surface. Figure~\ref{Fig4}(c) displays the similar temporal evolutions of the number of microcolonies from both the simulations and model Eq.\ref{eqFin}, showing agreement. 

\section*{Discussion}
What is the mechanism responsible for the different fraction of motile bacteria (i.e. heterogeneity) on the various surfaces? To gain further insights, we have submitted, after three days of experiments of growth, some samples to a high flux of deionized water and found that the amount $p_{rins}$ of remaining microcolonies is inversely correlated to the Young modulus (see Additional info). Since the two PDMS samples have the same surface energy, physico-chemical effects are not relevant and this shows that adhesion is enhanced by the softness of the surface. This basic result suggests an intuitive correlation between adhesion and the dynamical arrest observed on soft surfaces.

The model is built at the particle scale and its parameters stand for unknown processes at a molecular scale. While it is clear that the type IV pili play a key role in bacterial adhesion~\cite{Berne2018, Valentin2019, Dansuk2018, Rodesney2017}, their complex interactions with surfaces and with other bacteria deserve to be studied extensively with, for example, mutant of variable number of pili or surfaces of controlled structure. The description proposed here does not, however, rely upon explicit cell-substrate or cell-cell adhesion mechanisms.

We show that strong interactions with the surface provides a high level of heterogeneity. Such heterogeneity~\cite{Vissers2018} is key in maintaining the integrity of the biofilm of other bacteria~\cite{Horne2016}, in immune evasion~\cite{Harman2012} and in some metastatic process~\cite{delattre1}. This suggests that the relevance of heterogeneous phenotypes among a given strain may be a generic feature of active matter whose emergence shall be further investigated.

\paragraph*{Conclusions}
We have observed that microcolonies grow preferentially on soft surfaces. This is due to the increased adhesion, giving rise to a high proportion of non-motile bacteria which cannot escape microcolonies. The order of magnitude of the number of microcolony is well-described by a kinetic model with a correction for the proportion of motile bacteria. Understanding how adhesion strength and heterogeneous dynamics are linked could be useful to the implementation of new strategies for limiting the virulence of pathogens. 
\begin{acknowledgments}
The authors would like to thank F. Chauvat for useful discussions and A. M\'ejean for the kind help regarding the culture of the cyanobacteria.
\end{acknowledgments}

\appendix

\section*{Experimental details}
\subsection*{Materials}
The wild-type strain of the cyanobacterium \textit{Synechocystis} sp. PCC 6803 is cultured in a BG11 standard mineral medium. Cell suspensions are placed under a white light of intensity of 1.3 W.m$^{-2}$ for 7 days, and then kept in the dark for 24 hours. The suspensions are finally let two hours under ambient light until starting an experiment. This protocol results in a homogeneous cellular behaviour on glass~\cite{Lenotre}. 

The surfaces on which microcolony grow are glass (contact angle with water $\theta \sim 60^\circ$) and two different polydimethylsiloxanes (PDMS, Sylgard 184, Dow Corning) prepared with ratios of curing agent/polymeric base of 1/5 and 1/25, and for which $\theta \sim 100^\circ$.

The growth of microcolonies is studied in an ``open cell" made of a Petri dish covered with a lid into which a small $1$ mm hole is drilled, ensuring gaz exchange throughout the eight days of the experiment. First, samples of the different surfaces are placed together at the bottom the cell. The cell is then filled with the bacterial suspension, the lid is sealed and the system is placed inside an incubator with white light (1.3 W.m$^{-2}$). The imaging setup, based on a standard microscope, is placed in the incubator to record images of each surface. Images ($400 \times 500 \mu$m$^2$) are then processed as explained in the main text.

Because of convection, the diffusive dynamics of the bacteria cannot be characterized conveniently in the open cell. We then used a previous configuration in a closed cell as in~\cite{Lenotre}. The dynamics on the hard glass surface is studied on the bare glass slide while for the soft and intermediate surfaces, millimetric droplets of PDMS are deposited on the glass slide and let cure two days at room temperature (only one substrate at  a time can be studied with this setup). A droplet of suspension is deposited on the glass slide and trapped with a glass cavity. This cell is sealed with grease and contains a small volume of liquid, thus totally suppressing convection. After one hour at rest so that sedimentation is complete, the motion of the bacteria is video recorded and analysed by particle tracking~\cite{Lenotre}. 

\subsection*{Probing adhesion}
To get a rough estimate of the interactions between the bacteria and the surfaces, some open cells were thoroughly washed after three days of growth, by submitting the substrates to a high tangential flux of de-ionized water, thus detaching some microcolonies. Then, image analysis of the washed surfaces provides the number of remaining bacteria from which the proportion of attached cells  $p_{rins}$ is deduced. We found $p_{rins}=44$, $56$ and $63$~$\pm 5 \%$ for the glass, PDMS 1/5 and PDMS 1/25 respectively.  

\subsection*{Microcolonies formation and detection}
\label{A2}
\par In order to detect the microcolonies on a surface, we start from raw images to which we apply a bandpass filter with ImageJ. The background of each image is subtracted before inversion and binarisation. Microcolonies are detected on a size criterion: groups of pixel containing less than 80 units (30 $\mu$m$^2$, corresponding to an effective radius of 3.1 $\mu$m) are ignored. The number of remaining clusters defines the number of microcolonies. Data are averaged over four different images.

\subsection*{Computation of the average number of bacteria on each surface}
\label{A3}
\par We have determined the number of bacteria $n_{frame}$ on a given image by summing the floor function of the area of each detected cluster divided by the area of an average particle $\mathcal{A}_b=\pi r_b^2=15 \, \mu$m$^2$. If the floor function is less than 1 we set the value to 1. Spots smaller than 16 pixels ($\pi r_a^2$ = 6.2 $\mu$m$^2$) are considered as artefacts and are removed from the analysis. This is summed up by Eq.~\ref{eqnt}, where P bacteria of size $\mathcal{A}_{i} \geq 16$ pixels are detected.
\begin{equation}
\label{eqnt}
n_{frame}=\sum_{i=1}^P max \ \left[1,\textrm{floor} \left( \frac{\mathcal{A}_{i}}{\mathcal{A}_{b}}\right)\right]
\end{equation}
We have repeated this procedure on four different areas of each surface. The number of detected bacteria $n(t)$ is defined as the average value of $n_{frame}$ taken on these four images at time $t$.
\section*{Numerical simulations}
\label{A4}
\par Different types of motility are taken into account given the experimental proportion of motile bacteria $p_m$, and we consider that motile bacteria have the same behavior overall, as suggested by the experiments in the closed cell. 

We define by $n_i$ the number of cells for the simulation, randomly placed on a domain of size $\mathcal{S}=2.10^5 \mu$m$^{2}$ with periodic boundary conditions, resulting in a particle number density $\rho=n/\mathcal{S}$. A fraction $1-p_{m}$ of these cells are set non motile and thus unable to move or detach. The other ones perform random Brownian motion with an effective diffusion coefficient $D=0.05$ $\mu$m$^{2}$.s$^{-1}$, as in experiments. During the diffusion of a particular bacteria, a random number between 0 and 1 is computed if a neighbor is found at a distance $d<d_a$. If the random number is less than $p_{att}=\delta t/t_a$, then the cells bind to each other. The process is repeated for each particle found at a distance less than $d_a$. $\delta t=1$~s is the time step for the simulations, such that $\delta t \ll min(t_a,t_b)$. Motile bacteria can detach from their neighbors with a probability $p_{det}=\delta t/t_b$, and then restart their diffusive motion.

To account for cellular growth as in Fig. 4, the number of simulated bacteria $\rho(t)$ is imposed the experimental logistic equation. The integer part $\mathcal{N}(t)$ of $n(t)=\rho(t)\mathcal{S}$ is computed, and a new particle is created and randomly placed if $\mathcal{N}(t+\delta t) > \mathcal{N}(t)$. At the end of the simulation, we consider as ``microcolonies'' bacterial aggregates that contain at least two bacteria. Data shown in Fig. 4(c) are obtained with $d_a$=3 $\mu$m, $t_a$=450 s and $t_b$=100 s.

\end{document}